\documentclass[letterpaper,10pt,conference]{ieeeconf}
\usepackage{color}
\usepackage[dvipsnames]{xcolor}
\usepackage{amsmath}
\usepackage{amssymb}
\usepackage{graphicx}
\usepackage{bm}
\usepackage{url}
\usepackage[]{algorithm2e}
\usepackage{algpseudocode}
\usepackage{epsfig} 
\usepackage{color}
\usepackage[normalem]{ulem}
\usepackage{cancel}
\usepackage{fixltx2e}
\usepackage{array}
\usepackage{booktabs}

\makeatletter
\newcommand{\revisionhistory}[1]{%
\@ifundefined{showrevisionhistory}{\relax}{%
{#1}%
}%
}
\makeatother

\begin{document}


\title{\LARGE \bf Model Comparison of a Data-Driven \\and a Physical Model for Simulating HVAC Systems}
\author{Datong~Zhou$^*$, Qie~Hu$^*$, and~Claire~J.~Tomlin
\thanks{Datong Zhou is with the Department of Mechanical Engineering, University of California, Berkeley, USA. {\tt\footnotesize datong.zhou@berkeley.edu}}
\thanks{Qie Hu and Claire Tomlin are with the Department of Electrical Engineering and Computer Sciences, University of California, Berkeley, USA. \texttt{[qiehu, tomlin]@eecs.berkeley.edu}}
\thanks{This work is supported in part by NSF under CPS:ActionWebs (CNS-0931843) and CPS:FORCES (CNS-1239166).}
\thanks{Datong Zhou is funded by the Berkeley Fellowship for Graduate Study.}
\thanks{*These authors contributed equally.}}

\maketitle
\thispagestyle{empty}
\pagestyle{empty}

\begin{abstract}
Commercial buildings are responsible for a large fraction of energy consumption in developed countries, and therefore are targets of energy efficiency programs. Motivated by the large inherent thermal inertia of buildings, the power consumption can be flexibly scheduled without compromising occupant comfort. This temporal flexibility offers opportunities for the provision of frequency regulation to support grid stability. To realize energy savings and frequency regulation, it is of prime importance to identify a realistic model for the temperature dynamics of a building. We identify a low-dimensional data-driven model and a high-dimensional physics-based model for different spatial granularities and temporal seasons based on a case study of an entire floor of Sutardja Dai Hall, an office building on the University of California, Berkeley campus. A comparison of these contrasting models shows that, despite the higher forecasting accuracy of the physics-based model, both models perform almost equally well for energy efficient control. We conclude that the data-driven model is more amenable to controller design due to its low complexity, and could serve as a substitution for highly complex physics-based models with an insignificant loss of prediction accuracy for many applications. On the other hand, our physics-based approach is more suitable for modeling buildings with finer spatial granularities. 
\end{abstract}


%
%


\section{Introduction}
\label{sec:Introduction}
According to \cite{Perez-Lombard:2008aa}, residential and commercial buildings account for up to 40\% of the total electricity consumption in developed countries, with an upward trend. Heating, ventilation and air-conditioning (HVAC) systems are a major source of this consumption \cite{:aa}. 
Nevertheless, their power consumption can be flexibly scheduled without compromi-sing occupant comfort, due to the thermal capacity of buildings. As a result, HVAC systems have become the focal point of research, with the goal of utilizing this source of consumption flexibility. From the point of view of energy efficiency, researchers have studied optimization of building control in order to minimize power consumption \cite{Siroky:2011aa, Parisio:2014aa}.  
More recently, it has been proposed to engage buildings in supporting the supply quality of electricity and the grid stability, by participating in the regulation of electricity's frequency \cite{Balandat:2014aa, Lin:2015aa, Vrettos:2014aa, Baccino:2014aa}.

All of the above research activities are based on a valid mathematical model describing the thermal behavior of buildings. 
Traditionally, buildings have been modeled with high-dimensional physics-based models such as resistance-capacitance (RC) models \cite{Maasoumy:2014ab, Sun, David, Hao_multizone}, TRNSYS \cite{Duffy:2009aa} and EnergyPlus \cite{Zhao2013EP}. These models are motivated by the thermodynamics of the building and explicitly model the heat transfer between components of the buildings. The advantage of such models is their high granularity of temperature modeling, but a drawback is their high dimensionality which makes them computationally expensive. On the other hand, a new direction of research attempts to identify lower-dimensional, data-driven models, e.g. with Input-Output models \cite{Lin:2015aa} and semiparametric regression \cite{Aswani:2012aa}. The purpose is to alleviate the computational complexity in expense for coarser and less accurate temperature predictions.

A crucial question that arises within these two extremes is the extent to which the estimated temperature model is compatible with controller design. 
Take Model Predictive Control (MPC) for example, where the classical physics-based models require an MPC strategy to be solved online with high computational demand. 
Even then, the inherent bilinearity ensuing from the physics of the HVAC system often requires robustification, in the form of stochastic MPC formulations with chance constraints \cite{Ma:2012aa}. 
In contrast, regression-based models provide convenient difference equations that are easy and fast to use for MPC. 
A logical question to ask is how lean a model can be for a reasonable control application, without trading off too much 
accuracy and granularity of the temperature predictions.


To the best of our knowledge, data-driven and physical models for the identification of temperature evolution in commercial buildings have only been studied in isolation and on individual testbeds (e.g. \cite{Ma:2011aa}, \cite{Siroky:2011aa}, \cite{Lin:2015aa}, \cite{Qie}). The identified models are often validated using simulation data \cite{David} or experimental data collected under controlled environments, e.g. without occupants \cite{Lin:2015aa}. 
This makes a comparison of the performance between these two types of models, for a real building under normal operation, impossible due to their different nature. 

In this paper, we identify a data-driven model, using semiparametric regression, and adapt our previous physics-based model for the same building \cite{Qie}, using a one-year period of experimental data. 
We provide a quantitative comparison using various metrics, including open-loop prediction accuracy and closed-loop control strategies.
We show that, despite the higher accuracy of the complex physics-based model compared to the low-dimensional data-driven model, the optimal control strategy with respect to HVAC operation cost while maintaining the thermal comfort of occupants is almost identical for both systems. 
This indicates that the data-driven model provides enough accuracy for controller design, unless controllers that must maintain building temperatures more accurately or with finer spatial granularities are needed.
Finally, a qualitative analysis of the advantages and disadvantages of each type of model, together with their suitability for different applications is provided. 

The remainder of this paper is organized as follows: In Section \ref{sec:Preliminaries}, we describe the testbed and the data collected for our research. Section \ref{sec:Data_Driven_Model} presents the identification process for a purely data-driven model with semiparametric regression, followed by Section \ref{sec:Physics_Based_Model}, which details the procedure for identifying a physics-based model. Section \ref{sec:Comparison} will then compare the performance of the data-driven model and the physical model under different metrics. We conclude in Section \ref{sec:Conclusion} with a summary of our current and intended future work.

%


\section{Preliminaries}
\label{sec:Preliminaries}

\subsection{Testbed for System Identification}
\begin{figure}[hbtp]
\centering
\vspace*{-0.4cm}
\includegraphics[scale=0.30]{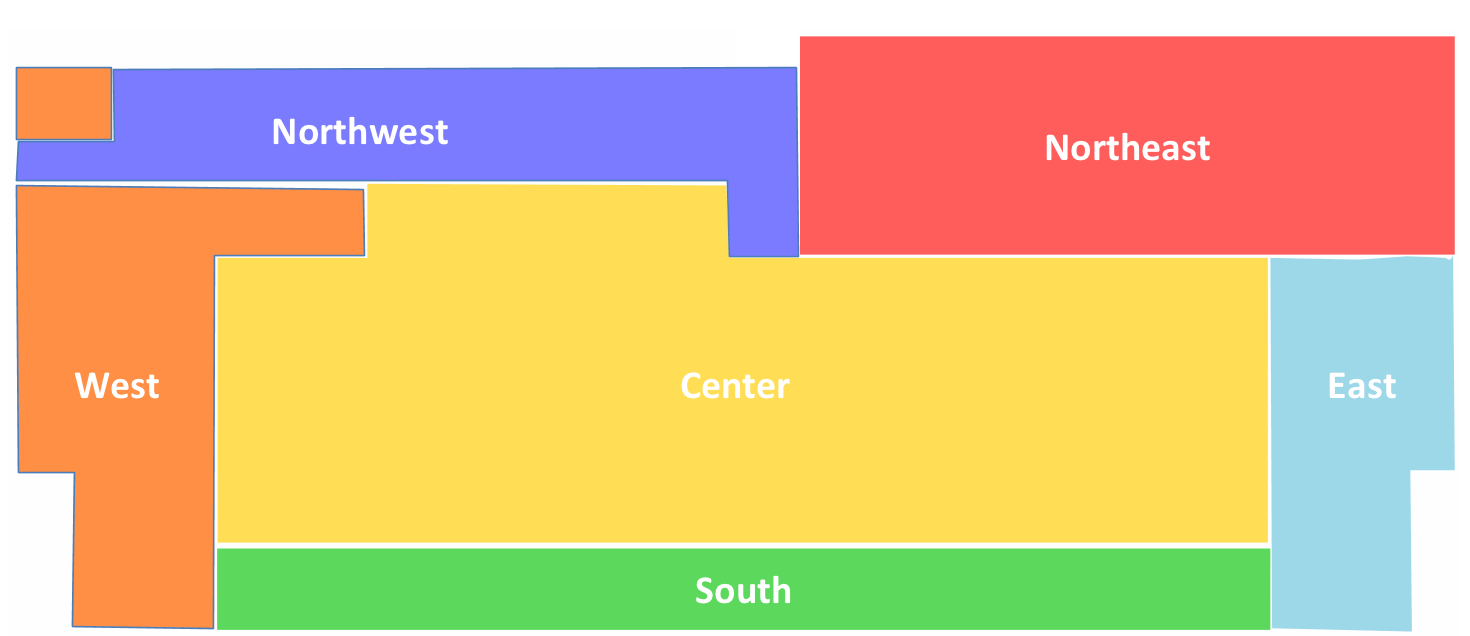}
\vspace*{-0.05cm}
\caption{Zones for the 4th Floor of Sutardja Dai Hall (SDH)}
\label{fig:floor_plan}
\vspace*{-0.45cm}
\end{figure}

We model the temperature evolution of the fourth floor of Sutardja Dai Hall (SDH), a building on the University of California, Berkeley campus. This floor contains offices for research staff and open workspaces for students, and is divided into six zones for modeling purposes (Figure \ref{fig:floor_plan}).

The building is equipped with a variable air volume (VAV) HVAC system that is common to 30\% of all U.S. commercial buildings \cite{VAV}. The system contains large supply fans which drive air through heat exchangers, cooling it down to a desired supply air temperature (SAT), and then distribute air to VAV boxes located throughout the building. There are 21 VAV boxes located on the fourth floor that govern the airflow to each room. In addition, the supply air may be reheated at the VAV box before entering the room. 


\subsection{Collection of Experimental Data}\label{sec:exp_data}
We collected 51 weeks of one-minute resolution temperature data for the six zones along with the airflow rates of the 21 VAV boxes, SAT and the outside air temperature from the \textit{simple Measurement and Actuation Profile} (sMAP). sMAP is a protocol that collects, stores and publishes time-series data from a wide variety of sensors \cite{:ac, Dawson-Haggerty:2012aa}. The hourly global horizontal solar radiation data recorded at a nearby weather station is obtained \cite{SolarRad}, from which the incidence solar radiation of the four geographic directions is calculated with the \texttt{PV\_LIB} toolbox \cite{:ab}. All collected data are down-sampled or interpolated, respectively, to 15 minute intervals. 

These 51 weeks of data span periods when the building was under normal operation as well as periods with excitation experiments. For accurate parameter identification, temperatures of neighboring zones should not have strong correlation \cite{Lin_multizone}. Our testbed is a regular office building in operation, thus forced response experiments were performed during Saturdays to (a) increase identifiability of the building model; (b) minimize effects due to occupancy on our data, and thus facilitate subsequent parameter identification; (c) minimize disturbance to building operation \cite{Qie}.


\subsection{Data Splitting}
Next, we define the seasons ``fall" (early September until mid December), ``winter" (mid December until late January), and ``spring" (late January until mid May) in order to account for different occupancy levels during the fall and spring semesters, and the winter break. After the weeks have been assigned to the seasons, a random portion of the data in each season (e.g. we chose 90\%) is defined as the training data, and the remaining weeks to be removed prior to the analysis are declared as the test set, which will be used to assess the accuracy of the optimal temperature model fitted on the training data.



%
%


\section{Data-Driven Model}
\label{sec:Data_Driven_Model}
We identify a difference equation for the temperature evolution with semiparametric regression, using the collected data from the 4th floor of SDH. Semiparametric regression in buildings has been proposed by \cite{Aswani:2012aa}, where the authors chose one week of data to model the temperature evolution including an exogenous heating load that captures the effect of occupancy, electric devices, VAV supply air temperature, outside air temperature, and solar radiation. We extend this approach by taking into account multiple weeks, which we separate into three seasons (fall, winter, spring) so as to characterize the different levels of the exogenous heating load for different temporal seasons. 

\subsection{Lumped Zone}
\label{sec:Lumped_Zone}
\subsubsection{Model Setup}
In order to facilitate analysis, the entire 4th floor of SDH is treated as a single zone, with the scalar temperature $x$ corresponding to the area-averaged zone temperatures and the input $u$ as the sum of the inflow of all 21 VAV boxes. This lumped model assumes a uniform temperature on the entire floor, $x$, and has been commonly used in literature \cite{Ma:2012aa, Oldewurtel:2010aa}. Then, the temperature evolution is assumed to have the following form:
\begin{equation}\label{eq:Temperature_evolution}
x(k+1) = a x(k) + b u(k) + c^\top v(k) + q_{\text{IG}}(k) + \epsilon(k),
\end{equation}
where $u$ denotes the total air inflow to the entire floor and $v := \left[v_\text{Ta}, v_\text{Ts}, v_\text{solE}, v_\text{solN}, v_\text{solS}, v_\text{solW} \right]^\top$ the vector of known disturbances that describe ambient air temperature, the HVAC system's supply air temperature and solar radiation from each of the four geographical directions.
In addition, $q_{\text{IG}}$ represents the internal gains due to occupancy and electric devices, and $\epsilon$ denotes independent and identically distributed zero mean noise with constant and finite variance which is conditionally independent of $x$, $u$, $v$, and $q_{\text{IG}}$.

\subsubsection{Smoothing of Time Series}
The unknown coefficients $a, b$, and $c$ are to be estimated using semiparametric regression \cite{Ruppert:2003aa, Hardle:2000aa}. The $q_{\text{IG}}$ term of Equation \eqref{eq:Temperature_evolution} is treated as a nonparametric term, so that \eqref{eq:Temperature_evolution} becomes a partially linear model. By taking conditional expectations on both sides of \eqref{eq:Temperature_evolution}, we obtain

\begin{equation}\label{eq:Conditional_Expectation}
\begin{aligned}
\hat{x}(k+1) &= a\hat{x}(k) + b\hat{u}(k) + c^\top\hat{v}(k) \\
& \quad + \mathbb{E}\left[ q_{\text{IG}}(k) \vert k \right] + \mathbb{E}\left[ \epsilon(k) \vert k \right],\\
\end{aligned}
\end{equation}
\noindent
where the conditional expectations $\hat{x}(\cdot) = \mathbb{E}\left[ x(\cdot) \vert \cdot \right]$, $\hat{u}(\cdot) = \mathbb{E}\left[ u(\cdot) \vert \cdot \right]$, and $\hat{v}(\cdot) = \mathbb{E}\left[ v(\cdot) \vert \cdot \right]$ are used.
Noting that $\mathbb{E}\left[ \epsilon(\cdot) \vert \cdot \right] = 0$ and assuming $\mathbb{E}\left[ q_{\text{IG}}(\cdot) \vert \cdot \right] = q_{\text{IG}}(\cdot)$, subtracting \eqref{eq:Conditional_Expectation} from \eqref{eq:Temperature_evolution} gives
\begin{equation}\label{eq:subtract}
\begin{split}
x(k+1) - \hat{x}(k+1) = a\left( x(k) - \hat{x}(k) \right) \\ + b\left( u(k) - \hat{u}(k) \right)
+ c^\top \left( v(k) - \hat{v}(k) \right) + \epsilon(k).
\end{split}
\end{equation}
The unknown internal gains term has been eliminated, and thus the coefficients $a, b, c$ in \eqref{eq:subtract} can be estimated with any regression method.

The conditional expectations $\hat{x}(\cdot), \hat{u}(\cdot)$ and $\hat{v}(\cdot)$ are obtained by smoothing the respective time series \cite{Aswani:2012aa}. We made use of locally weighted linear regression with a tricube weight function, where we use $k$-fold cross-validation to determine the bandwidth for regression.
The error measure used for in-sample estimates is the \textit{Root Mean Squared} (RMS) \textit{Error} between the measured temperatures $\bar{x}(k)$ and the model's predicted temperatures $x(k)$ over a time horizon of $N$ steps (e.g. we chose a 24 hour time horizon, $N = 96$):
\begin{equation}\label{eq:RMS}
\text{RMS error} = \left(\frac{1}{N}\textstyle\sum_{k=1}^N \left[\bar{x}(k) - x(k)\right]^2 \right)^{1/2}.
\end{equation}
\subsubsection{Bayesian Constrained Least Squares}
A main challenge in identifying the model is that commercial buildings are often insufficiently excited. Take SDH for example, whose room temperatures under regular operation only vary within a range of 2$^{\circ}$C and inflow of the single VAV boxes hardly vary at all. To overcome this, data collected during forced response experiments described in Section \ref{sec:exp_data} was used in training the model. To further compensate for the lack of excitation, a Bayesian regression method is used, which allows our prior knowledge of the building physics to be incorporated in the identification of coefficients. More specifically, Gaussian prior distributions are used for the coefficients $a$ and $b$, i.e., $a \sim \mathcal{N}( \mu_a, \Sigma_a)$ and $b \sim \mathcal{N}( \mu_b, \Sigma_b)$, where $\mathcal{N}( \mu, \Sigma)$ denotes a jointly Gaussian distribution with mean $\mu$ and covariance matrix $\Sigma$. In addition, $a$, $b$ and $c$ are constrained to be identical for the different seasons, since they model the underlying physics of the building which are assumed to be invariant throughout the year. Therefore the coefficient identification problem is formulated as follows:
\begin{equation}\label{eq:data_opt}
\begin{aligned}
(\hat{a}, \hat{b} , \hat{c}) = &\arg\min_{a, b, c}~\left(J_\text{f} + J_\text{w} + J_\text{s}\right) + \Vert \Sigma_a^{-1/2} (a - \mu_a) \Vert^2\\
& \quad\quad + \Vert \Sigma_b^{-1/2} (b - \mu_b)  \Vert^2 \\
\text{s.t.}~ J_i = &\Vert x_i(k+1) - \hat{x}_i(k+1) - a\left( x_i(k) - \hat{x}_i(k) \right)\\
& - b\left( u_i(k) - \hat{u}_i(k) \right) - {c}^\top\left( v_i(k) - \hat{v}_i(k) \right) \Vert ^ 2\\
& \text{for } i \in \lbrace \text{f}, \text{w}, \text{s} \rbrace, \\
& 0 < a < 1,~ b \leq 0,~ c \geq 0, \\
\end{aligned}
\end{equation}
where subscripts f, w, and s represent fall, winter and spring, respectively. The sign constraints on the parameters $b$ and $c$ translate into the fact that the temperature to be estimated positively correlates with all components in $v$ and negatively correlates with the VAV airflow. The range of $a$ is a consequence of Newton's Law of Cooling.
To find the effect of the VAV inflow on the 15-minute temperature evolution, we computed the 15-minute incremental decreases in temperature $\Delta x$ recorded during the excitation experiments. It is assumed that the large inflow $u$ dominates all other effects such that we can assume
\begin{equation}\label{eq:excitation_equation}
\Delta x = x(k+1) - x(k) = b\cdot u(k)
\end{equation}
for all $k$ during the excitation period. The estimated prior $\mu_b$ can then be isolated from \eqref{eq:excitation_equation}. The prior $\mu_a$ was set as the optimal $\hat{a}$ identified by \eqref{eq:data_opt} without the prior terms. The covariance matrices $\Sigma_a$ and $\Sigma_b$ were chosen subjectively. 
\subsubsection{Estimation of Internal Gains}
With the estimated coefficients $\hat{a}, \hat{b}, \hat{c}$ in hand, the internal gains $q_{\text{IG}}$ can be estimated by manipulating \eqref{eq:Conditional_Expectation}: 
\begin{equation}\label{eq:qig_estimation}
\hat{q}_{\text{IG}}(k) = \hat{x}(k+1) - \underbrace{\left(\hat{a}\hat{x}(k) + \hat{b}\hat{u}(k) + \hat{c}^\top \hat{v}(k)\right)}_{\mathbb{E}\left[x(k+1)\right]}.
\end{equation}
This can be interpreted as the difference between the smoothed temperature $\hat{x}(k+1)$ and the predicted expected temperature $\mathbb{E}\left[x(k+1)\right]$. With the estimated parameter coefficients being constant over the different seasons, a distinct function of internal gains is estimated for each season by averaging the estimated weekly gains for a given season.


\subsubsection{Results}
\begin{figure}[hbtp]
\centering
\vspace*{-0.0cm}
\hspace*{-0.6cm}
\includegraphics[scale=0.22]{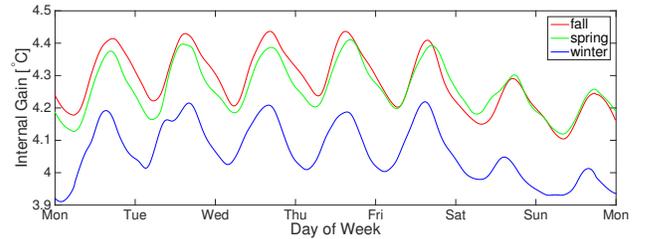}
\vspace*{-0.3cm}
\caption{Estimated Internal Gain $q_{\text{IG}}$ from the Data-Driven Model by Season, Lumped Case}
\vspace*{-0.3cm}
\label{fig:qig_comparison_plot}
\end{figure}

The estimated internal gains for each season, calculated with \eqref{eq:qig_estimation}, are shown in Figure \ref{fig:qig_comparison_plot}.
Observe that, for all three seasons, the internal gains exhibit a daily trend with local peaks around the late afternoon and local minima at night. Moreover, the amplitudes of the internal gains are considerably smaller during the weekends suggesting a lighter occupancy. It can further be seen that the magnitude of the internal gains is smallest for the winter season, which is in accordance with our intuition since most building occupants are absent during that period.


Lastly, since the Bayesian Constrained Least Squares algorithm \eqref{eq:data_opt} has identified a set of parameter estimates $\hat{a}, \hat{b}, \hat{c}$ valid for all three seasons to account for the time-invariant physics of the building, the temperature predictions are of the same nature for all three seasons. We thus conclude that the inherent differences between the seasonal temperature data are captured by the internal gains and can be compared between the seasons on a relative level.

The identified models for the different seasons found with \eqref{eq:data_opt} are
\begin{equation}\label{eq:lumped_zone_res}
\begin{aligned}
x(k+1) & = 0.80\cdot x(k) - 0.18\cdot u(k) \\
 & ~~ + \left[0.0019, 0.028, \mathbf{0} \right] v(k) + q_{\text{IG},i}(k) \\
 & =  0.80\cdot x(k) - 0.18\cdot u(k) \\
 & ~~ + 0.0019 \cdot v_{\text{Ta}}(k) + 0.028 \cdot v_{\text{Ts}}(k) + q_{\text{IG},i}(k) \\
 & \text{for } i \in \lbrace \text{f}, \text{w}, \text{s} \rbrace\\
\end{aligned}
\end{equation}
The estimated coefficients of $c$ corresponding to the solar radiation disturbances are very small ($< 10^{-6}$) compared to the other estimated coefficients. Since the temperatures are of the order $10^{\circ}$C, air inflow around 1 kg/s and solar radiation about 100 W/m$^2$, the effect of solar radiation on the room temperature is orders of magnitude less than that of other factors and hence can be neglected.

The average RMS prediction errors are 0.22$^{\circ}$C, 0.17$^{\circ}$C and 0.23$^{\circ}$C for fall, winter and spring respectively, showing that our model predicts the temperature reasonably well. 

\subsection{Individual Zones}\label{sec:Indiv_Zones}
\subsubsection{Model Setup}
Rather than approximating the entire 4th floor of SDH as a single zone, in this section, we identify a multivariate model that describes the thermodynamic behavior of each of the six individual zones:
\begin{equation}
\begin{aligned}\label{eq:temp_propagation_indiv}
x(k+1) &= A x(k) + B u(k) + C v(k) + q_{\text{IG},i}(k) \\
& ~~ \text{for } i \in \{ \text{f, w, s}\},
\end{aligned}
\end{equation}
where $x$, $q_{\text{IG},i} \in \mathbb{R}^6$, and the control input $u \in \mathbb{R}^6$ represent the temperatures, the internal gains of each zone, and the total air flow to each zone, respectively. In the lumped case, it was observed that solar radiation only had a negligible effect on the building's thermodynamics compared to the input and other disturbances, and thus we omit the solar radiation in the subsequent analysis: $v := \left[ v_\text{Ta}, v_\text{Ts} \right]^\top \in \mathbb{R}^2$.

Inspired by Newton's Law of Cooling, only adjacent zones influence each other's temperature, which defines the sparsity pattern of the coefficient matrices that are to be estimated. Hence 
\begin{equation}
A_{ij} = \begin{cases}
      \neq 0, & \text{if}\ i=j~\text{or}~(i,j)~ \text{adjacent}  \\
      0, & \text{otherwise.}
    \end{cases}
\end{equation}
The diagonal elements of $A$ denote autoregressive terms for zone temperatures, whereas non-diagonal elements describe the heat exchange between adjacent rooms. The matrix $B$ is diagonal by definition of $u$. The sparsity pattern of $C$ is found by physical adjacency of a respective zone to an exterior wall of a given geographic direction.

\subsubsection{Model Identification}
The procedure for the estimation of the parameter matrices $\hat{A}$, $\hat{B}$, $\hat{C}$, and the internal gains follows \eqref{eq:data_opt}, but with a modified choice of the (now matrix-valued) priors $\mu_a$ and $\mu_b$: $\mu_b$ and the diagonal entries of $\mu_a$ are obtained by scaling the corresponding priors from the lumped zone case in order to account for the thermal mass of the individual zones, which is smaller than in the lumped case. The off-diagonal elements of $\mu_a$, which represent the heat transfer between adjacent zones, were set to a value close to zero, according to our calculations with the heat transfer equation $\dot{q} = U\cdot A \cdot \Delta x$ and \cite{Koehler:2013aa}.
\subsubsection{Results}
\begin{figure}[hbtp]
\centering
\vspace*{-0.2cm}
\includegraphics[scale=0.46]{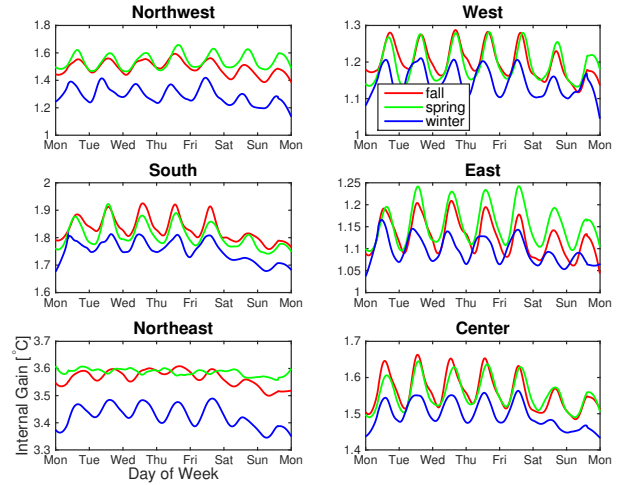}
\vspace*{-0.5cm}
\caption{Estimated Internal Gain $q_{\text{IG}}$ from the Data-Driven Model by Zone and Season, Individual Case}
\label{fig:qig_seasons_indiv}
\end{figure}

Figure \ref{fig:qig_seasons_indiv} shows the estimated internal gains for the three seasons fall, winter, and spring for the six single zones, computed with the smoothed time series \eqref{eq:qig_estimation}. It can be seen that the different zones exhibit different magnitudes of internal gains, with average values of the internal gains ranging between 1.0$^{\circ}$C and 3.6$^{\circ}$C for different zones and seasons. Similar to the lumped zone case (Figure \ref{fig:qig_comparison_plot}), daily peaks of the internal gains profiles can be recognized, with a slight decrease in magnitude on weekend days. The average prediction RMS error by zone and season are reported in Table \ref{tab:data_RMS_zones}.	

\begin{table}[hbtp]
\centering
\begin{tabular}{*8c}
\toprule
\multicolumn{8}{c}{Data-Driven Model} \\
\hline
Season & NW & W & S & E & NE & C & Mean \\ \hline
Fall & 0.98 & 0.61 & 0.28 & 0.42 & 0.28 & 0.36 & 0.488\\
Winter & 1.41 & 0.34 & 0.29 & 0.26 & 0.25 & 0.21 & 0.460\\
Spring & 0.56 & 0.25 & 0.31 & 0.71 & 0.17 & 0.34 & 0.390\\
\midrule
\midrule
\multicolumn{8}{c}{Physics-Based Model} \\
\hline
Season & NW & W & S & E & NE & C & Mean \\ \hline
Fall & 0.61 & 0.46 & 0.39 & 0.39 & 0.20 & 0.32 & 0.396\\
Winter & 0.55 & 0.39 & 0.34 & 0.32 & 0.18 & 0.24 & 0.338\\
Spring & 0.45 & 0.28 & 0.24 & 0.33 & 0.09 & 0.19 & 0.263\\
\bottomrule
\end{tabular}
\caption{RMS by Zone and Season for Data-Driven and Physics-Based Models}
\label{tab:data_RMS_zones}
\end{table}

%

%
%


\section{Physics-Based Model} \label{sec:Physics_Based_Model}

In this section, we describe the physics-based modeling approach proposed in \cite{Qie}, which is a Resistance-Capacitance (RC) model obtained using the Building Resistance-Capacitance Modeling (BRCM) MATLAB toolbox \cite{David}. 
A main advantage of this approach is that the resulting model has a small number of parameters, even for a complex multi-zone building; furthermore, these parameters have strong physical meaning, which aids in their identification. 

In this paper, we re-identify the building model using the same training dataset as used in Section \ref{sec:Data_Driven_Model}, and estimate distinct internal gains functions for different seasons.



\subsection{Model Setup}\label{sec:physics_model}

The physics-based building model has the following form \cite{Qie}:
\begin{subequations}\label{eq:physics_model}
\begin{align}
x(k+1) &= Ax(k)+B_v v(k) + B_\text{IG} f_\text{IG}(k) \label{eq:physics_model1} \\ \nonumber
	& \quad + \textstyle \sum_{i=1}^{21} \big( B_{xu_i} x(k) + B_{vu_i} v(k) \big) u_i(k) \label{eq:physics_model_2} \\
y(k) &= C x(k),
\end{align}
\end{subequations}
where the state vector $x \in \mathbb{R}^{289}$ represents temperatures of all building elements on the 4th floor and $y \in \mathbb{R}^6$ represents the average temperatures of the six zones shown in Figure \ref{fig:floor_plan}. $u \in \mathbb{R}^{21}$ denotes the air inflow rates, whose $i$-th element $u_i$ denotes the inflow rate from the $i$-th VAV box.
$v := [v_\text{Ta}, v_\text{Ts}]^\top$ is the disturbance vector, which captures known disturbances from ambient air temperature and the HVAC system's supply air temperature. 
Note that from our previous studies, heat gains due to solar radiation are orders of magnitude less than those caused by other disturbances and inputs and hence, are not included here. 
Finally, $f_\text{IG}(k) : \mathbb{N} \rightarrow \mathbb{R}^6$ captures internal gains in each of the six zones on the 4th floor and consists of two terms:
\begin{equation}\label{eq:fig}
f_\text{IG}(k) = f_\text{IG}^c + f_{\text{IG},m}^v(k) ~\text{for} ~ m \in \lbrace \text{f}, \text{w}, \text{s} \rbrace,
\end{equation}
where $f_\text{IG}^c$ is an unknown constant vector representing background heat gains due to idle appliances such as computers and printers. Functions $f_{\text{IG},m}^v(\cdot)$ for $m \in \lbrace \text{f}, \text{w}, \text{s} \rbrace$, are unknown nonparametric functions that capture the time-varying heat gain due to occupancy, equipment and other unmodeled uncertainties such as reheating at the VAV boxes in fall, winter and spring, respectively.
The system matrices $A$, $B_v$, $B_\text{IG}$, $B_{xu_i}$ and $B_{vu_i}$ are functions of tuning parameters: the window heat transmission coefficient ($U_\text{win}$), the convection coefficients of the interior wall ($\gamma_\text{IW}$), the exterior wall ($\gamma_\text{EW}$), the floor ($\gamma_\text{floor}$), and the ceiling ($\gamma_\text{ceil}$). 
Define $\gamma := \begin{bmatrix} U_\text{win}, \gamma_\text{IW}, \gamma_\text{EW}, \gamma_\text{floor}, \gamma_\text{ceil}, f_\text{IG}^{c\top} \end{bmatrix}^\top \in \mathbb{R}^{11}$, then to identify the physics-based model, we need to estimate the parameter vector $\gamma$ as well as the functions $f_{\text{IG},m}^v(\cdot)$.

Next, we describe our approach for identifying this model.

\subsection{Model Identification}\label{sec:physics_id}

For a fair comparison, the same data used to train and test the data-driven model is used to train and validate the physics-based model. 
The model identification process is performed in two steps: First, the subset of the training data collected during weekends is used to estimate the parameters, $\gamma$. Second, the nonparametric functions $f_{\text{IG},m}^v(\cdot)$ are estimated from the complete training dataset.

\vspace*{0.2cm}
\subsubsection{Parameter Estimation}
For parameter estimation purposes, we first set $f_{\text{IG},m}^v(\cdot) = 0$ during the weekend days, and evaluate them at a later point (Equations (\ref{Eq:ig(k-1)}) and (\ref{Eq:fixed_ig})). With $f_{\text{IG},m}^v(\cdot) = 0$, (\ref{eq:physics_model}) reduces to a purely parametric model:
\begin{equation}\label{eq:physics_parammodel}
\begin{aligned}
x(k+1) &= Ax(k)+B_v v(k) + B_\text{IG} f_\text{IG}^c\\
	& \quad + \textstyle \sum_{i=1}^{21} \big( B_{xu_i} x(k) + B_{vu_i} v(k) \big) u_i(k), \\
y(k) &= C x(k).
\end{aligned}
\end{equation}
The optimal model parameters are estimated by solving the following optimization problem:
\begin{equation}\label{eq:physics_opt}
\begin{aligned}
 \hat{\gamma} =&~\arg\min_{\gamma > 0}~\left(J_\text{f} + J_\text{w} + J_\text{s} \right) \\
\text{s.t.~~}
&J_m = \textstyle \sum_k \Vert y_m(k,\gamma) - \bar y_m(k) \Vert ^ 2 ~ \text{for} ~ m \in \lbrace \text{f}, \text{w}, \text{s} \rbrace\\
&y_m(k,\gamma) \text{~and~} x_m(k,\gamma) \text{~satisfy (\ref{eq:physics_parammodel}) with}\\
&x_m(0) = x_{\text{KF},m}(0)\\
&u_m(k) = \bar u_m(k), v_m(k) = \bar v_m(k)~\forall~k,\\
\end{aligned}
\end{equation}
\noindent
where $\bar u$, $\bar v$ and $\bar y$ denote the measured inputs, disturbances, and zone temperatures, respectively. In other words, we choose $\gamma$ such that, when the model is simulated with this set of parameter values and the measured inputs and disturbances, the sum of squared errors between the measured zone temperatures and the simulated temperatures is minimized.
The initial state $x_m(0)$ is required to simulate the model, however, not all states are measurable (the wall temperature for example is not), thus we estimate the initial states using a Kalman Filter $x_{\text{KF},m}(0)$, and set  $x_m(0) = x_{\text{KF},m}(0)$.
Furthermore, to compensate for the lack of sufficient excitation of the building, initial guesses for $\gamma$ that are physically plausible are chosen. The optimal parameter values are similar to those reported in \cite{Qie}, hence are not included here due to space limitations.


\vspace*{0.2cm}
\subsubsection{Estimation of $f_\text{IG}^v(\cdot)$ for Each Season}
Let $\mathcal{W}_m = \{1,2,\ldots,n_m\}$ denote the set of weeks in the training data for season $m$, and let  $f_{\text{IG},m,w}^v(\cdot)$ be an instance of the internal gains function $f_{\text{IG},m}^v(\cdot)$ estimated for week $w$ in $\mathcal{W}_m$. The optimal $\hat {f}_{\text{IG},m}^v(\cdot)$ is defined as the average of all estimates for a given season. 

More specifically, let $\tilde x(k)$ and $\tilde{y}(k)$ denote the predicted states and zone temperatures at time $k$, with $f_{\text{IG},m,w}^v(k-1) = 0$, i.e.,
\begin{equation}\label{Eq:xnoig_ynoig}
	\begin{aligned}
	\tilde x(k) & = Ax(k-1)+B_v v(k-1) + B_\text{IG} f^c_\text{IG} \\
		& \quad + \textstyle \sum_{i=1}^{21} \big( B_{xu_i} x(k-1) + B_{vu_i} v(k-1) \big) \\
		& \quad \cdot u_{i}(k-1), \\
	\tilde y & = C \tilde x(k).
	\end{aligned}
\end{equation} 
By noting $x(k) = \tilde{x}(k) + B_\text{IG} f^v_{\text{IG},m,w}(k-1)$, we can estimate $f^v_{\text{IG},m,w}(k-1)$ by solving the following set of linear equations using Ordinary Least Squares:
\begin{equation}\label{Eq:ig(k-1)}
(C B_\text{IG}) \cdot f_{\text{IG},m,w}^v(k-1) = \bar{y}(k) - \tilde{y}(k),
\end{equation}
where $\bar{y}(k)$ represents measured zone temperatures at time $k$.  
Finally, $\hat {f}_{\text{IG},m}^v(\cdot)$ is chosen as the average of all the estimates:
\begin{equation}\label{Eq:fixed_ig}
\hat {f}_{\text{IG},m}^v(k) = \frac{ \textstyle \sum_{w=1}^{n_m} f^v_{\text{IG},m,w}(k)}{n_m} \quad \forall~k.
\end{equation}
Therefore, the estimated function $\hat {f}_{\text{IG},m}^v(\cdot)$ takes into account the effect of hour of the day and day of the week on the internal gains.


\subsection{Results}\label{sec:physics_results}

\begin{figure}
	\centering
	\vspace*{-0.3cm}
	\includegraphics[scale=0.46]{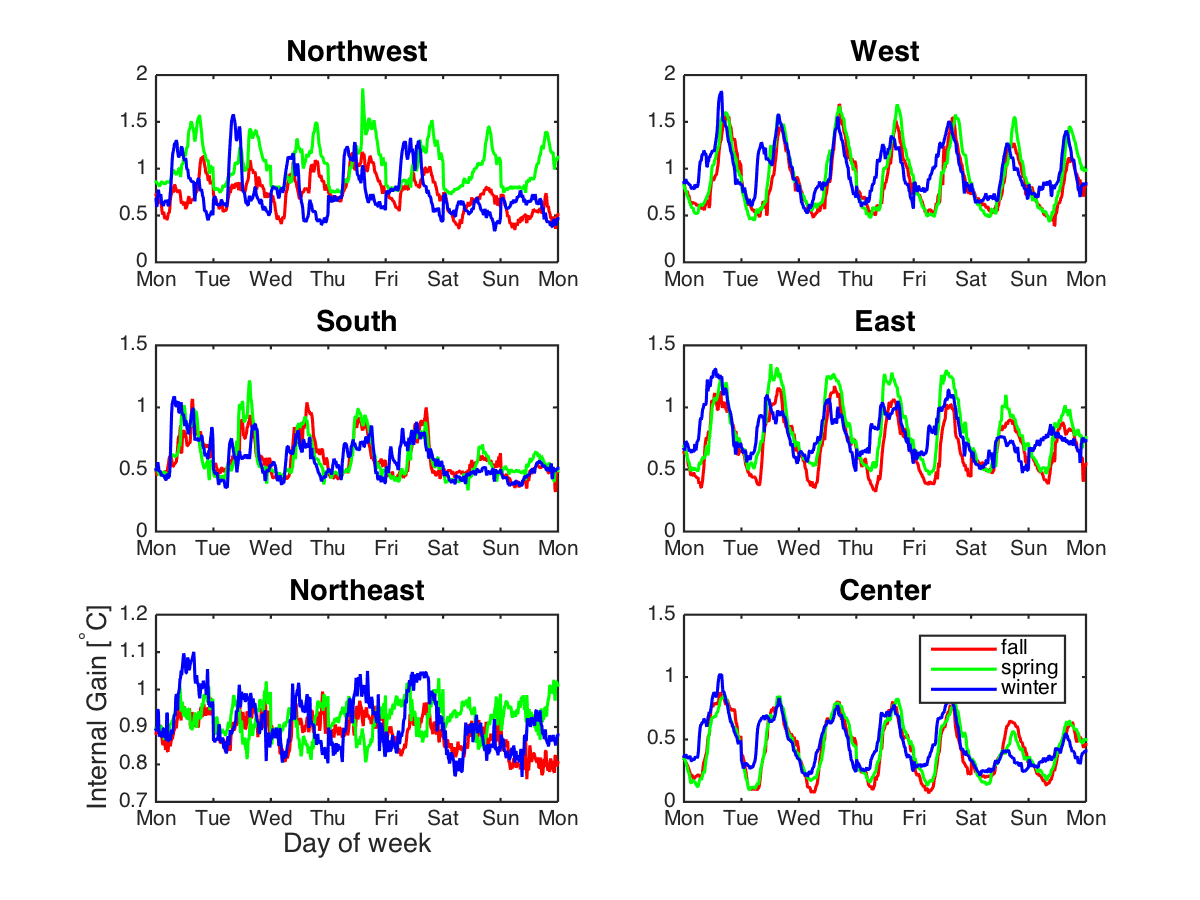}
	\vspace*{-0.2cm}
	\caption{Estimated Internal Gain $f_{\text{IG}}$ from the Physics-Based Model by Zone and Season}
	\vspace*{-0.5cm}
	\label{fig:physics_qig}
\end{figure}

The identified model is tested on holdout test weeks from different seasons. The average daily prediction RMS errors by zone and season are reported in Table \ref{tab:data_RMS_zones}. Figure \ref{fig:physics_qig} shows the estimated increase in zone temperatures due to internal gains for fall, winter and spring: $B_\text{IG}\cdot \big( f^c_\text{IG} + f^v_{\text{IG},m}(k) \big)$ for $m \in$ \{f, w, s\}, respectively. Similar average internal gains are observed for all zones and seasons. The zones that correspond to open workspaces and conference rooms (``West'', ``South'', ``East'' and ``Center'') show discernible daily peaks in their internal gains profiles with a slight decrease during weekends. Furthermore, there is little variation in the internal gains profiles across different seasons.

%
%


\section{Model Comparison}
\label{sec:Comparison}

\subsection{Prediction Accuracy}\label{sec:prediction_accuracy}
\begin{figure}
\centering
\vspace*{-0.3cm}
\includegraphics[scale=0.46]{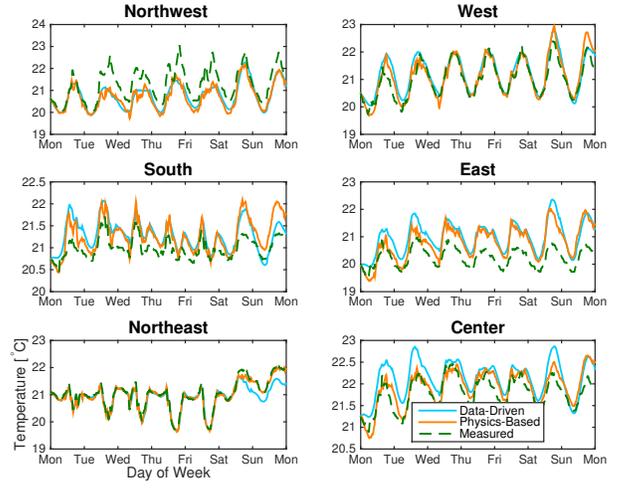}
\vspace*{-0.2cm}
\caption{Simulated Temperatures from the Data-Driven Model (blue), Physics-Based Model (orange) and Actual Temperatures (green)}
\vspace*{-0.5cm}
\label{fig:open_loop_trajectories}
\end{figure}

The high-dimensional physics-based model (Model B) is found to have a higher prediction accuracy compared to the low-dimensional data-driven model for the individual zones (Model A) presented in Section \ref{sec:Indiv_Zones}: According to Table \ref{tab:data_RMS_zones}, the mean RMS for Model B across zones is more than 0.1 degrees lower than for Model A. This is also illustrated in Figure \ref{fig:open_loop_trajectories}, which shows 7-day open-loop predictions of the temperature of a selected holdout test week in the spring period, simulated with both models instantiated once with an initial condition that matches the measured temperatures. The increase in RMS from Model B to Model A is notably larger in the zones ``East'' (0.38) and ``Center'' (0.15), compared to the other zones (0.11, $-$0.03, 0.07, and 0.08). 

\subsection{Energy Efficient Control}\label{sec:energy_efficient_control}
In this section, we compare the performance of Model A and Model B for the purpose of energy efficiency. We formulate an MPC problem to find the optimal control strategy that minimizes the cost of HVAC operation over the same week used in Figure \ref{fig:open_loop_trajectories}, while guaranteeing the temperature to stay within a comfort zone $[T_{\text{min}}, T_{\text{max}}]$, which we chose as $[20^\circ \text{C}, 22^\circ \text{C}]$ \cite{Hansen:2013aa}, and confining the control input to the physical limits of the HVAC system $[u_{\text{min}}, u_{\text{max}}]$. This problem is formulated as follows:
\begin{equation}\label{eq:ee_controller}
\begin{split}
\min_{u, \varepsilon}~&\sum_{k=1}^N u(k)^2 + \rho \Vert\varepsilon\Vert_2 \\
\text{s.t.}~& x(0) = \bar{x}(0) \\
& x(k+1) = \begin{cases}
      \eqref{eq:temp_propagation_indiv}, & \text{Model A}  \\
      \eqref{eq:physics_model1}, & \text{Model B}
    \end{cases} \\
& u_{\text{min}}-\varepsilon \leq u(k) \leq u_{\text{max}}+\varepsilon\qquad \forall k\in[0, N-1]\\
& \begin{cases}
      T_{\text{min}} \leq x(k) \leq T_{\text{max}}, & \text{Model A}  \\
      T_{\text{min}} \leq Cx(k) \leq T_{\text{max}}, & \text{Model B}~\eqref{eq:physics_model_2}
    \end{cases}~ \forall k\in[1, N]
\end{split}
\end{equation}
The temperature is initialized with the measured temperature $\bar{x}(0)$ at the beginning of the week-long simulation. We use soft constraints on the control input with a penalty parameter $\rho$ to ensure the feasibility of the problem. The penalty represents the cost of increasing the airflow beyond the operating limits (temporary shutdown or overuse, both of which are harmful to the system). 
To find the optimal control strategy, we make use of receding horizon control with a prediction horizon of three 15-minute time steps.

\begin{figure}
\centering
\vspace*{-0.4cm}
\includegraphics[scale=0.46]{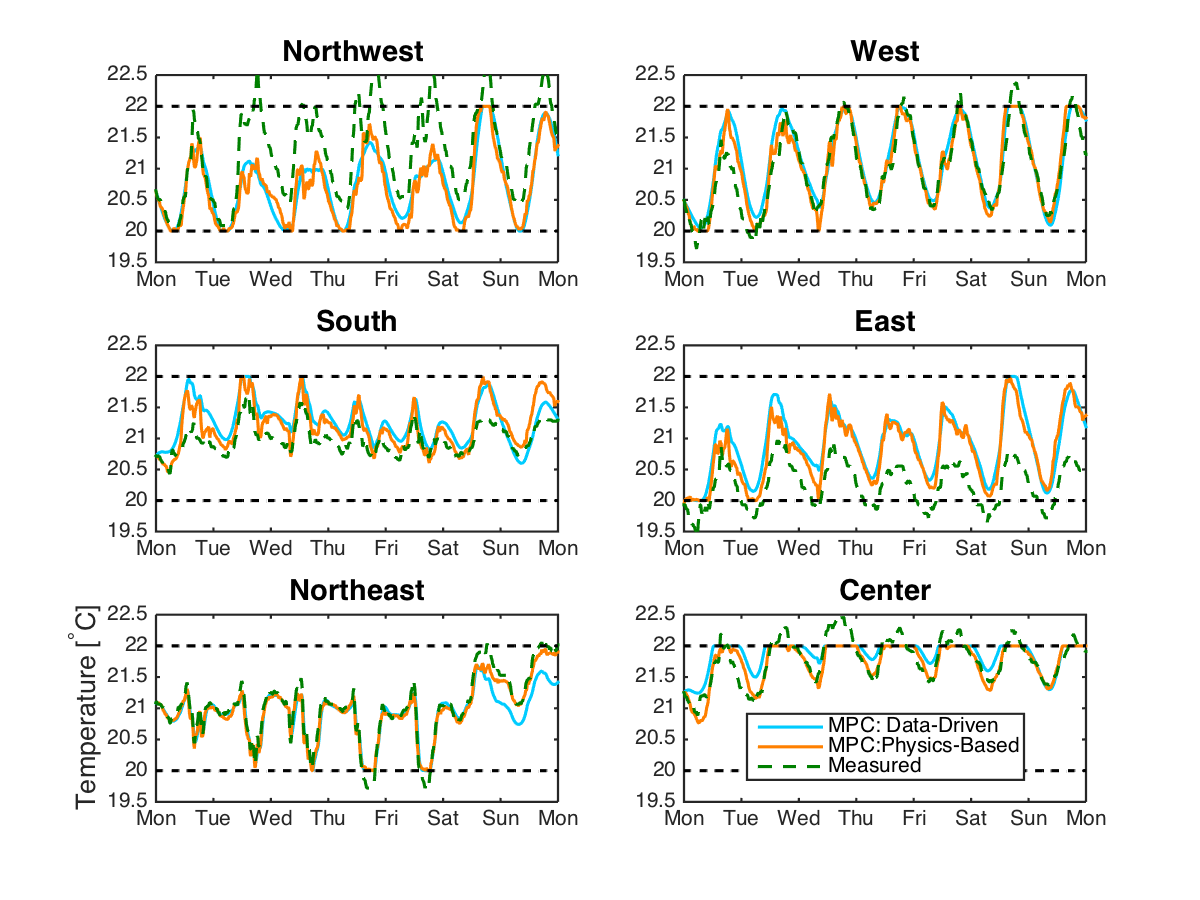}
\vspace*{-0.5cm}
\caption{Optimal Temperature for MPC with Data-Driven Model (blue), MPC with Physics-Based Model (orange) and Actual Temperature (green)}
\vspace*{-0.2cm}
\label{fig:MPC_comparison_temp}
\end{figure}

\begin{figure}[t]
\centering
\vspace*{-0.4cm}
\includegraphics[scale=0.46]{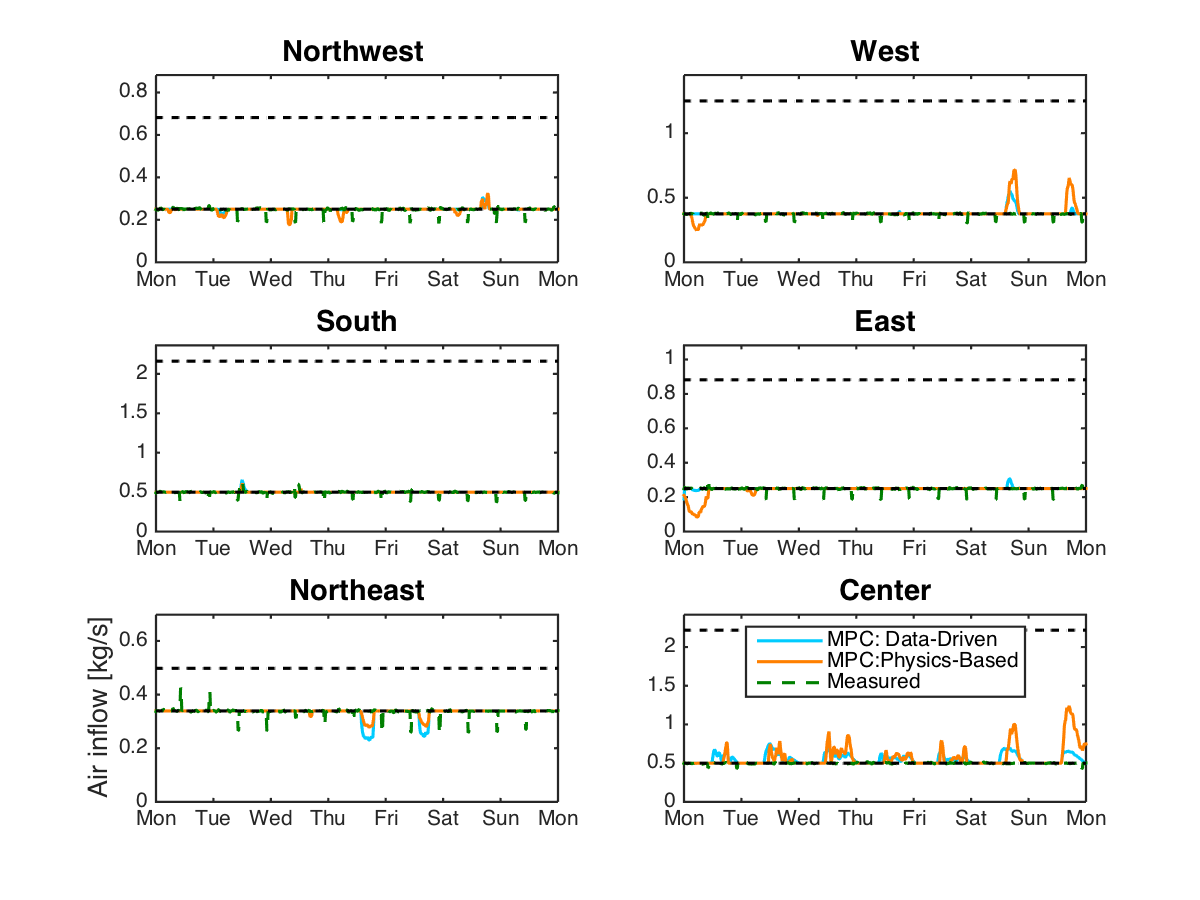}
\vspace*{-0.5cm}
\caption{Optimal Control Strategy for MPC with Data-Driven Model (blue), MPC with Physics-Based Model (orange) and Actual Input (green)}
\vspace*{-0.2cm}
\label{fig:MPC_comparison_flow}
\end{figure}

Figure \ref{fig:MPC_comparison_temp} shows the temperature trajectory computed by the energy efficient controller \eqref{eq:ee_controller} computed with both models A and B, together with the measured temperature as a reference. 
It can be seen that both control schemes are capable of maintaining the temperature within $[20^\circ, 22^\circ]$, with a control strategy that is of comparable cost (1,006 and 1,731 for Model A and Model B, respectively, where $\rho=100$), shown in Figure \ref{fig:MPC_comparison_flow}. An interesting observation is that the largest difference in the control strategies is detected in zones ``East'' and ``Center'', which show a larger increase in RMS from Model B to Model A.

The dips of the computed control trajectory $u$ below the black dashed lines represent violations of the physical limits needed to maintain the temperature in the narrow range $[20^\circ, 22^\circ]$. Furthermore, it is interesting to observe that variations in the control input do not impact the periodicity of the temperature qualitatively, which can be explained by the regularity of the identified internal gains.

These findings suggest that both models perform equally well in designing an energy efficient control strategy. However, computing this strategy for Model A was cheap ($<5$ minutes) compared to Model B ($\approx 20$ hours) on a 2 GHz Intel Core i7, 16 GB 1600 MHz DDR3 machine. Further, we note that in real-world applications, the MPC would use state feedback to initialize the temperature with sensor measurements at every time step, whereas in our simulation, it operates in an ``open loop'' fashion and hence propagates the estimation error with time. This, in essence, reduces the difference in the prediction quality by both controllers, since the RMS error is now to be evaluated on a much shorter prediction horizon, thereby further corroborating the finding of almost identical control schemes. For temperature-critical zones in which precise temperature estimations are needed, however, one might still want to choose the fine-level Model B for analysis.

\subsection{Qualitative Comparison}
We summarize the findings in the following:
\begin{itemize}
\item Model A is more \textit{amenable} to controller design due to its low dimensionality and hence considerably faster operation. This is of particular importance for \textit{scalability} considerations, since the computation time grows exponentially with the number of state variables, which renders Model B computationally intractable for online operation beyond a certain complexity. Indeed we observe that the computation for one step of \eqref{eq:ee_controller} exceeds 15 minutes $-$ the discretization time $-$ for a prediction horizon of five steps. Thus in frequency regulation, for instance, Model A's low dimensionality makes it suitable for reserve determination which must be computed for a time horizon of 24 hours.
\item Identifying Model A with semiparametric regression only relies on temperature and VAV airflow data and the physical adjacency of zones, in contrast to Model B, which requires knowledge of the BRCM toolbox and a large amount of geometry and construction data of the buildings, many of which are often unknown \cite{Qie}. Hence, more effort is required to train the model on new buildings for Model B.
\item The higher \textit{accuracy} of Model B proves useful for applications such as the control of temperature-critical zones and evaluation of controller performance through simulations, whereas Model A is preferably used for controller design and in applications where less emphasis is put on estimation errors, e.g. at night when building occupancy is low. 
\item Model A assumes a uniform temperature among zones, which often encompass several rooms, whereas Model B can provide estimates for the temperature of individual rooms in a given zone. The number of parameters of Model A increases rapidly with the model complexity, which coupled with insufficient excitation of the system makes it hard to emulate the higher spatial \textit{granularity} with Model A. 
\end{itemize}

\section{Conclusion}
\label{sec:Conclusion}

We identified state-space models for the thermal behavior of SDH with semiparametric regression and a physics-based model. The internal gains due to occupants and electric devices were identified for different spatial granularities and different temporal seasons. We found the high-dimensional physics-based model to yield lower estimation errors than the low-dimensional data-driven model due to the inclusion of analytical temperature models based on physical parameters of the building, therefore allowing for higher granularity in temperature predictions. Under an energy efficient MPC scheme, both models performed equally well, with the disadvantage of the physics-based model being computationally expensive due to its large number of states, which show an inherent bilinear relationship with inputs.

We note that the higher fidelity physics-based model should be used for controlling temperature-critical zones in buildings, since it provides higher granularity in addition to higher accuracy. The compact data-driven model, however, is a good alternative for devising a control strategy when less emphasis is put on estimation errors, e.g. at night when occupancy is low. In frequency regulation, the lower-dimensional data-driven model is more suitable for reserve determination as it requires planning over a longer time horizon, whereas the more accurate higher-dimensional physics-based model can be used in reserve provision to maintain the building temperature within comfort bounds and track the frequency regulation signal. Furthermore, while semiparametric regression can be easily applied on any building with a modest requirement of recorded data, the physics-based model requires detailed geometry and construction data about the building, which in practice is often subject to large inaccuracies, and therefore hard to obtain.

We are currently designing a control scheme suitable for frequency regulation in commercial buildings, based on the findings outlined in this paper. Further, we will validate the identified models by implementing these control schemes into the building operation system of SDH.

\section*{Acknowledgment}
We thank Anil Aswani for fruitful discussions.

\bibliographystyle{IEEEtran}
\bibliography{Bibliography}


\end{document}